\documentstyle[12pt]{article}
\newcommand{\be}{\begin{equation}}
\newcommand{\ee}{\end{equation}}
\newcommand{\bea}{\begin{eqnarray}}
\newcommand{\eea}{\end{eqnarray}}
\newcommand{\ket}{\rangle}
\newcommand{\bra}{\langle}
\newcommand{\bm}[1]{\mbox{\bf #1}}
\newcounter{fig}
\newcounter{crit}
\renewcommand{\thecrit}{\roman{crit}}
\begin{document}

\pagestyle{empty}
\begin{titlepage}

\title{$E2$ properties of nuclei far from stability\\
and the proton-halo problem of $^8$B}

\author{H. Nakada$^{(1,3)}$ and T. Otsuka$^{(2,3)}$\\
$^{(1)}$ {\small\em Department of Physics, Juntendo University,}\\
{\small\em Hiraga-gakuendai 1-1, Inba-mura, Inba-gun,
Chiba 270-16, Japan}\\
$^{(2)}$ {\small\em Department of Physics, Faculty of Science,
University of Tokyo,}\\
{\small\em Hongo 7-3-1, Bunkyo-ku, Tokyo 113, Japan}\\
$^{(3)}$ {\small\em Institute for Nuclear Theory,
University of Washington,}\\
{\small\em HN-12, Seattle, WA 98195, U. S. A.}}

\date{}

\maketitle
\thispagestyle{empty}

\noindent
{\small PACS numbers: 21.10.Ky, 23.20.--g, 21.60.Cs, 27.20.+n}
\\

\begin{abstract}
$E2$ properties of $A=6$--$10$ nuclei,
including those of nuclei far from stability,
are studied by a $(0+2)\hbar\omega$ shell-model calculation
which includes $E2$ core-polarization effects explicitly.
The quadrupole moments and the $E2$ transition strengths
in $A=6$--$10$ nuclei are described quite well
by the present calculation.
This result indicates that the relatively large value
of the quadrupole moment of $^8$B
can be understood without introducing the proton-halo in $^8$B.
An interesting effect of the $2\hbar\omega$ core-polarization
is found for effective charges
used in the $0\hbar\omega$ shell model;
although isoscalar effective-charges are almost constant
as a function of nucleus,
appreciable variations are needed for isovector effective-charges
which play important roles in nuclei with high isospin-values.
\end{abstract}

\end{titlepage}

\setcounter{page}{1}
\pagestyle{plain}
\section{Introduction\label{sec:intro}}

The structure of light neutron-rich nuclei
has recently attracted much interest.
A good example is the $^{11}$Li nucleus\cite{ref:Li11},
which shows the exotic feature of the neutron-halo,
owing to loosely bound neutrons.
Research on light proton-rich nuclei is also in progress.
One then comes up with a question
whether or not the proton-halo exists despite the Coulomb barrier.
Recently Minamisono {\it et al.} succeeded in a precise measurement
of the quadrupole moment of $^8$B\cite{ref:Minamisono}.
They pointed out\cite{ref:Minamisono,ref:Kitagawa} that
the quadrupole moment of $^8$B is considerably larger
than the shell-model prediction.
They claimed the existence of proton-halo\cite{ref:Minamisono}
based on the analysis of Ref.\cite{ref:Kitagawa}.
This analysis, however, appears to be model-dependent,
as will be discussed later.

On the other hand, the observed interaction radius of $^8$B
shows no enhancement,
compared with those of surrounding nuclei\cite{ref:Tanihata}.
In contrast to the case of $^{11}$Li,
this datum undoubtedly contradicts
with the proton-halo hypothesis\cite{ref:Minamisono,ref:Kitagawa}
of $^8$B.
Thus, a serious conflict arises between Refs.\cite{ref:Minamisono}
and \cite{ref:Tanihata}.

In this paper, we shall consider $E2$ properties of light nuclei,
from a more general perspective
including nuclei far from stability.
While the isoscalar degrees of freedom dominate
the $E2$ properties of low-lying states of light stable nuclei,
the isovector ones become important
in unstable nuclei with higher isospin.
This work will shed light on a new aspect
in the structure of light unstable nuclei,
namely the $E2$ properties.

As will be shown later,
it is difficult to avoid
in the $0\hbar\omega$ shell-model calculation
ambiguities arising from effective charges
as well as from single-particle wavefunctions
({\it i.e.}, single-particle matrix-elements).
It is of much importance to remove such ambiguities
in effective charges.
For this purpose, a shell-model calculation
with explicit inclusion of the $2\hbar\omega$ excitation
is very useful,
because this excitation is the major origin of the effective charges
and therefore the effective charges are
almost equal to the bare charges
in the $(0+2)\hbar\omega$ shell model.
We investigate the $E2$ properties
in terms of the $(0+2)\hbar\omega$ shell-model calculation
by using the interaction introduced
by Wolters {\it et al.}\cite{ref:Wolters}.
Although several problems have been pointed out
with respect to this interaction\cite{ref:Millener},
it has a certain advantage in investigating $E2$ properties.
We shall also discuss
some effects of the $2\hbar\omega$ configuration
on the isoscalar and isovector effective-charges.

\section{$(0+2)\hbar\omega$ shell-model calculation\label{sec:0+2}}

The interaction of Wolters {\it et al.}\cite{ref:Wolters}
has been determined for the $(0+2)\hbar\omega$ shell-model calculation,
so as to fit the experimental energies,
including the binding energies, of $A=4$--$16$ nuclei.
In this paper, we shall primarily discuss results obtained by a
$(0+2)\hbar\omega$ shell-model calculation with this interaction.
The harmonic-oscillator basis is employed
with a constant $\hbar\omega$ for all of those nuclei.
Since the effective interaction is given
in terms of the relative coordinates
as the values of the Talmi integrals,
it is free from the spurious center-of-mass motion.
The energy levels of $^8$Li and $^8$B, which are mirror nuclei,
are shown in Fig.\ref{fig:energy}, in comparison with the data.
The Coulomb energies are subtracted
in the same manner as in Ref.\cite{ref:Wolters}.
Note that the isospin symmetry is maintained
in the present calculation.
The result of the $0\hbar\omega$ shell-model calculation
with the Cohen-Kurath (8--16)TBME interaction\cite{ref:CK}
is also displayed in Fig.\ref{fig:energy}.
The ground-state energy in this $0\hbar\omega$ result
is adjusted to the experimental one.
It is clear that the present $(0+2)\hbar\omega$ calculation
reproduces the data quite well:
the agreement is better than the $0\hbar\omega$
result with the Cohen-Kurath interaction.

Although in the $(0+2)\hbar\omega$ calculation
of Ref.\cite{ref:Wolters} $Q$-moments were adjusted
to the data of Ref.\cite{ref:Selove2},
we shall not adopt those parameters because of the following reasons.
While we investigate both $Q$-moments and $B(E2)$ values,
no transition was discussed in Ref.\cite{ref:Wolters}.
The experimental value of $Q(^8$Li$)$,
which is particularly important in discussing $Q(^8$B$)$,
is confirmed recently\cite{ref:Minamisono,ref:moment}.
This value is considerably larger than that shown
in Refs.\cite{ref:Selove2,ref:Selove}.
The parameters in Ref.\cite{ref:Wolters}
do not reproduce the presently confirmed value of $Q(^8$Li$)$.

We restrict ourselves to $A=6$--$10$ nuclei,
so as to keep the parameters almost constant.
Because the $E2$ core-polarization effect is included significantly
in the $(0+2)\hbar\omega$ space,
it is expected that one can reproduce the $E2$ properties
to a great extent with the bare charges.
The calculated $Q$-moments and $B(E2)$'s with the bare charges
are shown in Table~\ref{tab:E2}, in comparison with the data.
The $0\hbar\omega$ results displayed in Table~\ref{tab:E2}
will be discussed later.
It is found that, by this $(0+2)\hbar\omega$ calculation,
the $Q$-moments are already reproduced
to an appreciable extent with the bare charges.
If we introduce a small value of the isoscalar effective-charge
correction, $\delta e_{\rm IS}=0.05e$
({\it i.e.}, $e_\pi^{\rm eff}=1.05e$ and $e_\nu^{\rm eff}=0.05e$),
the agreement with the data in $A=6$--$8$ is improved further
as shown in Table 1.
With this parameter we come to a very good agreement in $Q(^8$Li$)$.
It is found that $Q(^8$B$)$ is reproduced
by the same parameter within reasonable accuracy ($\sim 8\%$).
Thus, one cannot claim that the $Q$-moment of $^8$B
is anomalously enhanced.
The structure of $^8$B can be understood within the shell model,
as the mirror nucleus of $^8$Li.
Because there is no evidence for halo in $^8$Li,
the proton-halo hypothesis for $^8$B appears to be very unlikely.
Although the $8\%$ difference in $Q(^8$B$)$
may originate from some halo-like structure
({\it i.e.}, slowly damping tail),
it should occur with a much reduced amplitude
compared with $^{11}$Li or $^{11}$Be\cite{ref:Li11,ref:Tanihata}.

As is discussed in Appendix,
the tail behavior of wavefunctions is connected
with separation energies,
although this connection is not taken into account
in most shell-model calculations
including the present one.
It is clear that the proton distribution of $^8$B is damped
more slowly in the radial direction than that of $^8$Li,
corresponding to the smaller separation energy.
The slower damping of the proton density
could lead to a stronger influence on the $Q$-moment.
Such an effect, on the other hand,
is not contained in the present calculation.
As discussed above, this might be a reason why we still have
a slight underestimation of $Q(^8$B$)$
with $\delta e_{\rm IS}=0.05e$,
by which we can reproduce $Q(^8$Li$)$ very well.
It should be noticed, however,
that the amplitude of the tail part of the wavefunction
is not connected with the separation energy,
as pointed out in Appendix.
The results of Ref.\cite{ref:Tanihata}
and the present work suggest consistently
that the amplitude of the slowly damping part
in the proton distribution of $^8$B
is too small to be regarded as proton-halo.
Namely, its influence on radius and $Q$-moment
does not appear to be so significant.
It is emphasized that,
although we have introduced a single parameter $\delta e_{\rm IS}$,
the value of this parameter is quite small
and therefore the resultant ambiguity becomes much less
than that arising from the effective charges needed for the
$0\hbar\omega$ space.
The present small value of $\delta e_{\rm IS}$
may come from the configurations
which are still outside the present shell-model space, and/or
from a mesonic effect.

A recent experiment\cite{ref:Li8E2} indicates
that the $B(E2; 2_1^+\rightarrow 1_1^+)$ value of $^8$Li
is quite large.
This datum appears to be far beyond
what can be explained by the present calculation,
or by the cluster-model calculation of Ref.\cite{ref:Baye}.

It is interesting that the $Q$-moments and $B(E2)$'s
of $A=9$ and $10$ nuclei
are reproduced quite well with the bare charges,
while $\delta e=0.05e$ is preferable in $A=6$--$8$.
There could be a slight change in the core-polarization effect
which originates in the excitations to higher $\hbar\omega$ space.

The oscillator length $b$ is derived
from the $\hbar\omega$ value fixed in the effective hamiltonian,
in the present calculation of $E2$ properties.
On the other hand, for computing charge radii and $Q$-moments
Wolters {\it et al.} adopted a smaller $b$-parameter
in Ref.\cite{ref:Wolters},
which is inconsistent with the $\hbar\omega$ value
for calculating energy levels.
This happened because Wolters {\it et al.}
tried to adjust the $b$-parameter
to the measured radii\cite{ref:radius}
of the whole region of $A=4$--$16$ nuclei.
However, as long as we restrict ourselves to $A=6$--$10$ nuclei,
the $b$-parameter adopted in Ref.\cite{ref:Wolters}
appears to be too small
to reproduce the charge radii of these nuclei.

The $b$-parameter used in the present work
reproduces the charge radius of  $^6$Li.
The approximate constancy of the radii
over the $p$-shell nuclei\cite{ref:radius}, however,
cannot be reproduced.
This is a common problem with the usual shell model,
especially when the harmonic-oscillator
single-particle wavefunctions are used.
If we consider the $0\hbar\omega$ space,
the radius must increase
with the mass number,
even with a constant $b$ value,
owing to the center-of-mass correction\cite{ref:Hees}.
This tendency does not change essentially
in the present $(0+2)\hbar\omega$ calculation,
whereas the experimental values increase more slowly.
There is an approach in which the $b$-parameter is determined
by the variation with a Skyrme-type interaction
for each nucleus\cite{ref:Skyrme}.
A nearly constant value of $b$ was then obtained for $A=6$--$12$,
implying that the above problem remains.
We shall not discuss the radii in further detail.
It is basically beyond the scope
of the usual shell-model calculations,
including the present one, to reproduce the radii.

\section{Effect of $2\hbar\omega$ configuration\label{sec:2hw}}

We shall consider the following ratio of the matrix elements,
\be R_T (i\rightarrow f) \equiv
{{\bra f | {\cal O}^{(T)} | i \ket}
  \over{\bra f | P_{0\hbar\omega} {\cal O}^{(T)} P_{0\hbar\omega}
   | i \ket \left/
  \sqrt{\bra f | P_{0\hbar\omega} | f \ket \bra i
   | P_{0\hbar\omega} | i \ket}\right. }},
{}~~(T=0,1) \label{eq:R} \ee
where the numerator is obtained
from the present $(0+2)\hbar\omega$ shell-model calculation,
$P_{0\hbar\omega}$ in the denominator represents
the projection operator onto the $0\hbar\omega$ space
({\it i.e.}, the space consisting
only of the $p$-shell configuration),
and $T=0$ ($T=1$) denotes the ratio
with respect to the isoscalar (isovector) operator.
Since $E2$ properties are under discussion,
we consider the case of ${\cal O}=r^2 Y^{(2)}$.
Apart from the possible changes
in the single-particle wavefunctions
between the $0\hbar\omega$ and the $(0+2)\hbar\omega$ calculations
mentioned below,
$R_T$ expresses, to a certain extent, the relative ratio
of the effective charge for the $0\hbar\omega$ calculation
over the bare charge,
because these effective-charges incorporate
the isoscalar and isovector $2\hbar\omega$ core-polarization effects.
The present single-particle wavefunctions
are determined so as to reproduce the energy levels
in the $(0+2)\hbar\omega$ space.
As a result, the $p$-shell single-particle wavefunctions
do not necessarily correspond
to the ones suitable
for the usual $0\hbar\omega$ shell-model calculation.
The overlap between these two sorts of $p$-shell wavefunctions,
however, is expected to be fairly large,
and therefore $R_T$ will serve
as a good measure of the effective charges
for the usual $0\hbar\omega$ shell-model calculation.

Figure~\ref{fig:effch} shows the $R_T$ values
for the $E2$ properties of $A=6$--$10$ nuclei.
For isovector matrix-elements,
$R_{T=1}$ is not shown for the matrix elements between $T=0$ states,
because it is indefinite.
We do not show the $R_{T=1}$ values
also when the isovector matrix-element is
less than 15\% of the isoscalar one.
Note that the isovector matrix-element reaches
40\% of the isoscalar one at maximum.
The $R_T$ values for $B(E2;0_2^+ \rightarrow 2_1^+)$ of $^{10}$Be
are omitted in Fig.\ref{fig:effch},
since the $0_2^+$ state is highly dominated
by the $2\hbar\omega$ configuration (97\%) in the present calculation.

It is remarkable that $R_{T=0}$ is fairly constant for $A=6$--$10$.
Most of the $R_{T=0}$ values are in the 0.9--1.2 range,
which leads to
$e^{\rm eff}_{\rm IS}\equiv {1\over 2}(e^{\rm eff}_\pi
+ e^{\rm eff}_\nu)=0.45e$--$0.6e$
for the $0\hbar\omega$ space.
We point out that $R_{T=0}$ for $^8$Li (or $^8$B)
is not extraordinary,
compared with their surrounding nuclei.
Since there is no enhancement in the $R_{T=0}$ value
for $Q(^8$Li$)$ and $Q(^8$B$)$,
a strong quadrupole deformation
is not likely to occur in $^8$B.
We here state that
the probability of the $2\hbar\omega$ configuration
shows no increase in the ground-state wavefunctions
of $^8$Li and $^8$B.
This result showing no notable increase
of the $2\hbar\omega$ mixing in $^8$Li
is consistent with the result
of a Hartree-Fock shell-model calculation in Ref.\cite{ref:HF+SM}.

The $R_{T=1}$ value shown in Fig.\ref{fig:effch}
fluctuates considerably.
The $R_{T=1}$ value changes from 0.9--1.5 for $A=7$ and $8$
to 0.0--0.5 for $A=9$ and $10$.
The $R_{T=1}$ values for the $Q$-moments of $^8$Li and $^8$B
are larger than those for the other nuclei,
which accounts for the relatively large difference
between $Q(^8$B$)$ and $Q(^8$Li$)$.
However, the $R_T$'s for $^8$Li and $^8$B still stay
within the range of its fluctuation.

The present systematic study suggests
that the isovector core-polarization effect changes
from nucleus to nucleus in lighter mass-region,
probably because the nuclear system is not large enough
and therefore the mean-field is less developed.
The $R_T$ values shown here indicate that
the $2\hbar\omega$ core-polarization effect
on isovector effective-charges
seems to have a significant nucleus-dependence.
Such variation of the core-polarization effect, however,
cannot be taken into consideration
within the $0\hbar\omega$ configuration space,
as far as constant (or almost constant) effective-charges are used.
The present extension of the model space
has an remarkable advantage
to take into account this variation of the core-polarization effect.

\section{Discussion\label{sec:discuss}}

\subsection{Assessments of proton-halo hypothesis
\label{subsec:proton-halo}}

It has been shown that $Q(^8$B$)$ can be understood
without introducing a rather exotic feature such as proton-halo,
contrary to the analysis in Refs.\cite{ref:Minamisono,ref:Kitagawa}.
Being consistent with
the interaction-radius measurement\cite{ref:Tanihata},
the proton-halo in $^8$B is not likely to exist.
At this stage we should reconsider the soundness
of the proton-halo hypothesis
in Refs.\cite{ref:Minamisono,ref:Kitagawa}.

The proton separation energy ($S_p$) of $^8$B
is very small ($\sim 0.14$MeV).
This was probably one of the basic motivations
for the proton-halo interpretation
in Refs.\cite{ref:Minamisono,ref:Kitagawa}.
It is known that the tail form of the $^8$B wavefunction
is connected with $S_p$,
as is discussed in Appendix for a more general case.
We should keep in mind, however,
that the separation energies
do not fix the amplitude of the tail part,
which is denoted by $\xi$ in Eq.(\ref{eq:asymp-n}).
In usual cases, this amplitude is considered to be small
enough to be neglected.
Only the halo nuclei,
in which
the nucleon occupation number in the tail region
is of order of magnitude one,
have appreciable contributions of the tail part
to various physical quantities.

In Ref.\cite{ref:Kitagawa},
Kitagawa and Sagawa compared two results
with different sets of single-particle wavefunctions.
One is obtained from the harmonic-oscillator potential,
while a Woods-Saxon potential is assumed in the other.
They applied the Cohen-Kurath shell-model density-matrix\cite{ref:CK}
to both cases.
Their procedure to determine the single-particle wavefunctions
from the Woods-Saxon potential
is explained in the following.
The ground-state of $^8$B was expanded
in the products of the $^7$Be ($^7$B) `core' states
and last proton (neutron).
For the $^7$Be$+$$p$ channel, this expansion is expressed as
\be \psi(\mbox{$^8$B};{\em g.s.}) =
 \sum_{i,j} c^p_{i,j} [\psi(\mbox{$^7$Be};i) \varphi_p(j)] ,
  \label{eq:cfp-p} \ee
where $\psi(\mbox{$^7$Be};i)$ denotes
the $i$-th eigenstate of $^7$Be,
$\varphi_p(j)$ the single-particle orbit $j$ of the last proton,
and $c^p_{i,j}$ stands for the spectroscopic amplitude.
The single-particle wavefunctions of the last proton were fixed
by the observed $S_p$ and the Woods-Saxon potential as shown below.
Note that the wave function $\varphi_p(j)$ is determined by this
method not only for the tail part but also for the inner part,
and is used for calculating physical observables.
The proton single-particle energies, $\epsilon_p(j)$, are determined
by the condition
\be E(\mbox{$^8$B};{\em g.s.}) = E(\mbox{$^7$Be};i) + \epsilon_p(j) ,
 \label{eq:weak-coupl} \ee
for each set of $i$ and $j$ in Eq.(\ref{eq:cfp-p}).
Note that $E(\mbox{$^8$B};{\em g.s.})$
and $E(\mbox{$^7$Be};i)$
are obtained from observed energies of the relevant states
of $^8$B and $^7$Be.
The depth of the Woods-Saxon potential was varied
for each combination of $i$ and $j$ separately,
so that the proton single-particle energy
should be adjusted to this value.
Since the proton separation energy is defined
by $S_p(\mbox{$^8$B}) = E(\mbox{$^7$Be};{\em g.s.})
- E(\mbox{$^8$B};{\em g.s.})$,
the value of $\epsilon_p(j)$ was fitted to $-S_p(\mbox{$^8$B})$,
when the state $i$ refers to the ground state
in Eq.(\ref{eq:weak-coupl}).
Associated with the small $S_p$,
this configuration led to the proton-halo.
It is noticed that the wavefunction of the last proton,
as well as the Woods-Saxon potential depth,
depend on the $^7$Be-core states in Eq.(\ref{eq:cfp-p}).
It was claimed in Refs.\cite{ref:Minamisono,ref:Kitagawa}
that this adjusted Woods-Saxon approach can reproduce $Q(^8$B$)$
with a set of effective charges similar to those of heavier nuclei.

The `adjusted Woods-Saxon' prescription described above
was originally developed by Millener {\it et al.}\cite{ref:Be11M}
in order to reproduce the tail form of the total wavefunction.
We point out here that,
through the condition of Eq.(\ref{eq:weak-coupl}),
the following two approximations are made in determining $\varphi_p(j)$
by the `adjusted Woods-Saxon' approach;
(a) The residual interaction
between the $^7$Be core and the last proton is ignored
(likewise the interaction between $^7$B and the last neutron).
(b) The coupling among different configurations was ignored.
In other words, off-diagonal matrix-elements of the nuclear force
are neglected among different configurations
in Eq.(\ref{eq:weak-coupl}).
The separation like Eq.(\ref{eq:weak-coupl})
could be valid when the quantum number of the state
is owed by the last proton.
This requires for the core state to be $0^+$.
Thus the `adjusted Woods-Saxon' manipulation may work better
for a loosely bound particle decoupled from its $0^+$ inert core.
The states discussed in Ref.\cite{ref:Be11M}
seem to be the cases of this kind.
On the other hand, several problems occur in $^8$B.

We list the problems on the proton-halo hypothesis
in Refs.\cite{ref:Minamisono,ref:Kitagawa} below.
\begin{list}{(\thecrit)}{\usecounter{crit}
\setlength{\leftmargin}{0cm}\setlength{\itemindent}{3.2em}}
\item As is stated in Section~\ref{sec:intro},
no enhancement of the interaction radius is observed
in $^8$B\cite{ref:Tanihata}.
\label{crit:radius}
\item The lowest $2^+$, $1^+$ and $3^+$ levels of $^8$B
correspond quite well to those of $^8$Li\cite{ref:Selove},
thus indicating good isospin symmetry.
Note that those isobaric analog states are also observed
as excited states of $^8$Be\cite{ref:Selove}.
Since the neutron-halo is not expected in $^8$Li,
the proton-halo hypothesis for the $^8$B ground-state
would destroy these isospin multiplets to a certain extent.
\label{crit:isospin}
\item In $^8$B, we have three protons in the $p$-shell,
as far as we work within the $0\hbar\omega$ shell-model space.
In the theoretical analysis
of Refs.\cite{ref:Minamisono,ref:Kitagawa},
Kitagawa and Sagawa seem to have distinguished
the orbits for the last proton
from those for the other two valence protons.
Hence, when the $^7$Be-core stays in the ground state
in the expansion of Eq.(\ref{eq:cfp-p}),
they assumed a loosely bound orbit for the last proton
and deeply bound orbits for the other two valence protons.
The antisymmetrization among the three valence protons, however,
was not treated correctly.
This can lead to erroneous results.
\label{crit:antisym}
\item The weak-coupling assumption stated above was fundamental
for the proton-halo hypothesis in Ref.\cite{ref:Kitagawa}.
It should, however, be tested carefully whether this assumption
is valid in $^8$B or not.
We shall estimate the coupling effects ignored
in Ref.\cite{ref:Kitagawa},
which are referred to as (a) and (b) in the preceding discussion,
by applying the $0\hbar\omega$ shell-model calculation
with the Cohen-Kurath interaction.
We obtain for the sum of the correlation energies,
$E(\mbox{$^7$Be};{\em g.s.}) + \epsilon_p(0p_{3/2})
- E(\mbox{$^8$B};{\em g.s.}) \sim 2$MeV.
This value seems too large to be ignored,
compared with $S_n(\mbox{$^8$Li})-S_p(\mbox{$^8$B}) \sim 2$MeV.
\label{crit:weak-coupling}
\item It was assumed in the theoretical interpretation
of Refs.\cite{ref:Minamisono,ref:Kitagawa}
that the effective charges should be the same
among the $A=8$, $11$ and $17$ nuclei.
Apart from the possible arbitrariness in their selection of nuclei,
the constant effective-charge assumption has to
be used with extreme care
to draw the proton-halo conclusion in the light mass region.
Among such light nuclei,
a considerable mass-number dependence of the effective charges
might be possible,
as we discussed in Section~\ref{sec:2hw}.
It can also be questioned whether or not
the effective charges should be common
among halo orbits and normal orbits.
\label{crit:effch}
\item The shell-model density-matrix
calculated with the Cohen-Kurath interaction\cite{ref:CK}
was employed in Ref.\cite{ref:Kitagawa}.
The Cohen-Kurath interaction, however,
is adjusted to the levels of non-halo nuclei.
It is not consistent to apply
those density-matrices to the halo orbits.
The influence of the halo on the density-matrix
should not be ignored,
in the case that the halo causes significant changes
of physical observables.
\label{crit:int}
\item The excited $1^+$ and $3^+$ states of $^8$B
are considered to have a certain similarity in structure
to the ground $2^+$ state,
since these three states are regarded
as the $(0p_{3/2})^4$, $T=1$ multiplet,
as a zeroth-order approximation within the $jj$-coupling scheme.
It is desirable, therefore, that the structure of the three states
should be described in a consistent manner.
However, if we apply the `adjusted Woods-Saxon' method
of Ref.\cite{ref:Kitagawa} to the excited states,
the following problem may occur.
The $3^+$ state is observed
2.2MeV above the threshold for proton emission.
Hence, Eq.(\ref{eq:weak-coupl}) leads to
$\epsilon_p\simeq 2.2$MeV for the $3^+$ state,
when the state $i$ refers to the ground state
in Eq.(\ref{eq:weak-coupl}) as in the $1^+$ state.
On the contrary, height of a Coulomb barrier is about 1MeV,
if we calculate it by using the Woods-Saxon potential
with the parameters of Ref.\cite{ref:BM1}.
The higher $\epsilon_p$ than the barrier makes the state
impossible to survive even as a resonance state.
\label{crit:weak-coupling2}
\item Let us consider the term in Eq.(\ref{eq:cfp-p})
with the $^7$Be state being the ground state.
Both the proton $0p_{3/2}$ and $0p_{1/2}$ orbits
can produce $2^+$ states by coupling to the $^7$Be ground state.
If the single-particle energies of the last proton
are determined from Eq.(\ref{eq:weak-coupl})
as in Ref.\cite{ref:Kitagawa},
we obtain $\epsilon_p(0p_{3/2})=\epsilon_p(0p_{1/2})$.
This implies that the $L\cdot S$-splitting
was ignored in Ref.\cite{ref:Kitagawa}.
\label{crit:LS}
\item The $\hbar\omega$ value adopted
for the harmonic-oscillator wavefunctions in Ref.\cite{ref:Kitagawa}
is questionable.
Although this value (and the corresponding $b$-parameter)
was important for the shell-model estimate of $Q(^8$B$)$,
the $\hbar\omega$ value was fixed
by the systematics among much heavier nuclei.
We shall return to this point in Subsection~\ref{subsec:b}.
\label{crit:hw}
\end{list}

As has been mentioned at point (\ref{crit:antisym}),
the correct antisymmetrization among nucleons is important.
It is not easy to carry out the full antisymmetrization
among constituent nucleons
and simultaneously reproduce the correct tail form
of the $^7$Be$+$$p$ channel.
In Sections~\ref{sec:0+2} and \ref{sec:2hw} we have been staying
within the shell model,
abandoning for the time being the reproduction of the tail behavior.
The shell model describes the nuclear surface region pretty well,
while in usual cases the tail region is beyond the scope of the model.
The quadrupole moment under discussion
is dominated by the surface region
for normal nuclei.
If we were not capable of reproducing the $Q$-moment
within the shell model,
it would suggest the possibility of halo.

\subsection{Reconsideration of the $b$-parameter
in $0\hbar\omega$ shell model\label{subsec:b}}

As mentioned at point (\ref{crit:hw}) in the previous subsection,
the harmonic-oscillator wavefunctions in Ref.\cite{ref:Kitagawa}
would hardly reproduce the properties of the nuclei around $A=8$.
We shall re-examine this point.

The Cohen-Kurath interaction\cite{ref:CK} has been frequently used
for the $0\hbar\omega$ shell-model calculation in the $p$-shell.
The discussion in Refs.\cite{ref:Minamisono,ref:Kitagawa}
was also based on this standard interaction.
We now consider the shell-model wavefunction
obtained from the Cohen-Kurath (8--16)TBME interaction.
In order to fix the oscillator length $b$,
we use the observed matter radius,
in contrast to Ref.\cite{ref:Kitagawa}.
The rms matter radii of $^8$B and $^8$Li
reported in Ref.\cite{ref:Tanihata}
lead to $b\simeq 1.77 {\rm fm}$,
after the center-of-mass correction.
This value of the $b$-parameter differs significantly
from the one adopted in Ref.\cite{ref:Kitagawa},
{\it i.e.}, $1.60{\rm fm}$.

The $Q$-moments and $B(E2)$ values in $A=6$--$10$ nuclei
computed with $b\simeq 1.77 {\rm fm}$ in the $0\hbar\omega$ space
with the Cohen-Kurath interaction
are shown in Table~\ref{tab:E2},
together with the $(0+2)\hbar\omega$ results
and the experimental data.
This $0\hbar\omega$ calculation
contains a single parameter $\delta e_{\rm IS}=0.5e$,
which leads to $e_{\rm IS}^{\rm eff}=1.0e$.
The agreement with the data seems to be rather good,
indicating the validity of the present isoscalar effective-charge.
It is found that $Q(^8$Li$)$ is reproduced well,
while $Q(^8$B$)$ is underestimated considerably.

If data of $Q$-moments in a set of mirror nuclei are available,
it is possible to determine the effective charges
for those nuclei from the data,
as was done in Ref.\cite{ref:Kitagawa}.
By applying this procedure to $^8$Li and $^8$B,
together with the present value $b=1.77 {\rm fm}$,
we obtain $e_\pi^{\rm eff} = 2.1e$ and $e_\nu^{\rm eff} = 0.3e$.
This set of effective charges is less surprising
than those obtained from the smaller value of
$b$ in Ref.\cite{ref:Kitagawa}.

The isoscalar effective-charge comes to $e_{\rm IS}^{\rm eff}=1.2e$,
whereas $e_{\rm IS}^{\rm eff}=1.6e$ in Ref.\cite{ref:Kitagawa}
for the harmonic-oscillator wavefunctions.
The analysis in Ref.\cite{ref:Kitagawa}
for the $A=11$ and $17$ nuclei
led to $e_{\rm IS}^{\rm eff}=0.9e$
and $e_{\rm IS}^{\rm eff}=1.0e$, respectively.
Compared with these values,
$e_{\rm IS}^{\rm eff}=1.2e$ for the $A=8$ nuclei is not anomalous.

Through the above discussions we recognize
that it is quite difficult to reach a conclusion
when dealing with such light nuclei,
as far as we restrict ourselves to the $0\hbar\omega$ space,
since both the effective charges
and the single-particle wavefunctions
give rise to ambiguities.
The calculated $Q$-moment is influenced by these ambiguities.

We further comment upon the analysis
using the single-particle wavefunctions
obtained from the ordinary Woods-Saxon potential.
If the Woods-Saxon potential
with the parameters in Ref.\cite{ref:BM1} is adopted,
the data of $Q(^8$Li$)$ and $Q(^8$B$)$ will lead to
$e_\pi^{\rm eff} = 2.3e$ and $e_\nu^{\rm eff} = 0.3e$.
If we vary some parameters
so as to fit the $\sqrt{\bra r^2 \ket}$ data,
quite similar effective-charges are required
to the harmonic-oscillator case
giving the same $\sqrt{\bra r^2 \ket}$.

\subsection{Explanation of the difference
between $Q(^8$Li$)$ and $Q(^8$B$)$\label{subsec:Q-dif}}

In this subsection we shall discuss
why $Q(^8$B$)$ is substantially larger than $Q(^8$Li$)$,
in an intuitive way within the $0\hbar\omega$ shell model.
Suppose that $Q$-moments are dominated
by isoscalar degrees of freedom,
equal values are obtained between mirror nuclei.
The $Q$-moments of $^{11}$B and $^{11}$C are so close,
exhibiting an example of this kind.
On the other hand, in experiment,
$Q(^8$B$)$ is about twice larger than $Q(^8$Li$)$.
Within the $jj$-coupling scheme without a residual interaction,
the $0p_{3/2}$ orbit is partly occupied
while the $0p_{1/2}$ is unoccupied,
either for protons and neutrons.
However, once we switch the residual interaction on,
it is much easier in $^8$Li for neutrons
to excite to $0p_{1/2}$ than for protons,
owing to the excess in number.
The inspection of the density matrix
and the single-particle matrix-elements
indicates that the excitation/de-excitation
of a nucleon between $0p_{3/2}$ and $0p_{1/2}$ yields
the principal contribution to the $Q$-moment.
It follows that $Q(^8$Li$)$ is dominated
by the neutron degrees of freedom,
unless $e_\nu^{\rm eff}$ value is not too small.
Similarly $Q(^8$B$)$ is governed by the proton degrees of freedom.
Consequently, $Q(^8$Li$)$ and $Q(^8$B$)$ reflect
the neutron and proton effective-charges, respectively,
giving rise to an enhanced sensitivity to the isovector charge.
Note that this mechanism does not work for the $A=11$ nuclei,
primarily because of the smaller difference
between proton and neutron number.

\subsection{Sensitivity of $Q(^8$B$)$ to interaction
\label{subsec:sensitivity}}

The shell-model interaction of Ref.\cite{ref:Wolters}
adopted for the present $(0+2)\hbar \omega$ calculation
has several problems,
as has been argued in Ref.\cite{ref:Millener};
the radial excitation is a primary problem in the argument.
Furthermore, the present wavefunctions do not necessarily provide us
with reasonable nuclear radii, as mentioned earlier.
Although we do not discuss these points here,
it should not be overlooked
how sensitive the present result of $Q(^8$B$)$
is to the choice of interaction.

It is known that,
since the quadrupole part dominates
the proton-neutron correlation in low-lying states,
the energy levels are sensitive
to the $Q_\pi \cdot Q_\nu$ component of a two-body interaction,
where
\be Q_\rho = \sum_{k\in\rho} r_k^2 Y^{(2)}(\hat{\bm r}_k)
 ~~ (\rho=\pi,\nu), \ee
with $k$ being the label of each nucleon
and $\hat{\bm r}_k = {\bm r}_k/r_k$.
This $Q_\pi \cdot Q_\nu$ part plays a significant role,
at the same time,
in describing $E2$ properties.
Hence the low-lying energy levels and the $E2$ properties
are not independent of each other.
Although effective charges may give rise to an ambiguity
in calculating the $E2$ properties,
this ambiguity will hardly influence
when the effective-charge corrections ($\delta e$'s) are small.
In such cases the $E2$ properties are dominated by $Q_\pi$.

The $E2$ properties of low-lying states depend predominantly
on angular correlations of nucleons in the nuclear surface region.
The low-lying spectrum is also dominated
by the surface degrees of freedom,
because the low-lying excited states are obtained usually
by exciting one to a few nucleons around the surface
from the ground state.
The $Q_\pi \cdot Q_\nu$ component of the interaction
carries an important part of this excitation.
On the other hand, the radial (or monopole) excitation
is much more dependent on the interior region
(and excitations from this region)
than the $E2$ properties and the low-lying spectra.
In other words, this excitation is much of the volume character.
Owing to this aspect,
the low-lying spectra and $E2$ properties
can be treated separately, to a certain extent,
from the problem of the radial excitations.

We shall now move to the problem of the radius.
This problem is more general and is not characteristic
to the present scheme, as stated earlier.  On the other hand,
the change of the radius from a nucleus to another
is treated independently of energies and $E2$ properties
in most shell-model calculations.
This is reasonable probably
because energies and $E2$ properties are more sensitive
to angular correlations in the surface region.
As an example, the $A$-dependence of the interaction
and that of the single-particle wavefunction
are not connected to each other
in the successful shell-model calculation
for the $sd$-shell \cite{ref:WB}.
A more precise description of the radius
seems to be a difficult task in a general term,
and the goal is rather far.
Note that the nuclear radius is also contributed
by the interior region to a considerable extent.
For a phenomenological interaction like the present one,
the problems of radial modes (radius and radial excitation)
may be separated from energy levels and E2 properties.

It has been shown in Ref.\cite{ref:Wolters}
that the low-lying energy levels around $A=8$ are reproduced
by the interaction of Wolters {\it et al.}
This interaction should contain the $Q_\pi \cdot Q_\nu$ part
in an effective manner.
The resultant good agreement
between calculated and experimental energy levels
over several nuclei suggests
that the effect of $Q_\pi \cdot Q_\nu$ is incorporated properly.
The $E2$ properties are also reproduced in the low-lying states
with only a small effective-charge correction,
by the present $(0+2)\hbar\omega$ calculation.
It therefore turns out that the present interaction is applicable
to the investigation of $E2$ properties of the $A\sim 8$ nuclei,
while there remain open issues in radii and radial excitations.
This consequence is in accordance with the general discussion
in the preceding paragraphs.

In the present case, the agreement in $Q(^8$Li$)$ ensures
the amount of $Q_\pi$ in $^8$Li,
which is exactly equal to the amount of $Q_\nu$ in $^8$B.
Since $Q_\pi \cdot Q_\nu$ is tested by the energy levels,
it is expected that the present $(0+2)\hbar\omega$ calculation
of the $Q(^8$B$)$ value is plausible.
Thus one can anticipate that
the present result with respect to $Q(^8$B$)$
will not be varied, for instance,
by a future tuning of the interaction,
as far as the interaction reproduces
both the energy levels and other related $E2$ properties.

\section{Summary\label{sec:summary}}

As long as we try to describe the $E2$ properties of light nuclei
within the $0\hbar\omega$ shell model,
the ambiguities arising from single-particle wavefunctions
and effective charges
are inevitable.
On the other hand,
the $Q$-moments and $B(E2)$ values of $A=6$--$10$ nuclei,
including $Q(^8$B$)$,
are reproduced within the present $(0+2)\hbar\omega$ shell model.
It is noticed that the $(0+2)\hbar\omega$ approach
is much less ambiguous
in fixing the values of effective charges than the calculation
restricted to the $0\hbar\omega$ space.
This $(0+2)\hbar\omega$ calculation suggests
that the proton-halo in $^8$B is very unlikely.
The observed $Q(^8$B$)$ value is not an evidence for proton-halo.
This result is consistent with the data
on the interaction radius\cite{ref:Tanihata}.
Even though the proton density distribution in $^8$B
may be somewhat spread,
its amplitude seems to be too small to be regarded as proton-halo.
The expectation value of the proton number in the tail region
will be much less than unity,
although such a reduced tail may still produce interesting effects
in astrophysical issues\cite{ref:deform}.
Though the present discussions have been based
on the interaction of Wolters {\it et al.}\cite{ref:Wolters}
and this interaction has several problems\cite{ref:Millener},
the conclusion on the $E2$ properties
are not very sensitive to the choice of the interaction.

The effect of the $2\hbar\omega$ configuration on $E2$
is also discussed.
The $2\hbar\omega$ contribution is customarily incorporated
into the $0\hbar\omega$ calculation by effective charges.
Indeed, the $R_T$ values defined in Eq.(\ref{eq:R}) are rather stable
for isoscalar matrix-elements.
However, those for isovector matrix-elements
are largely $A$-dependent.
This result suggests an intriguing aspect that,
although the isoscalar effective-charges
for the $0\hbar\omega$ space could be almost constant,
appreciable variation is required for the isovector effective-charges,
which play more important roles in nuclei far from stability
with higher isospin.

\section*{Acknowledgments}

The authors are grateful to Prof. T. Minamisono,
Prof. H. Sagawa and Dr. D. J. Millener
for valuable discussions.
The authors appreciate the hospitality
of the Institute for Nuclear Theory at the University of Washington,
and thank Prof. A. Gelberg for careful reading of the manuscript.
This work is supported partly by Research Center for Nuclear Physics,
Osaka University, as RCNP Computational Nuclear Physics Project
(Project No. 92-B-03).
This work is supported in part by Grant-in-Aid
for General Scientific Research (No. 04804012),
by the Ministry of Education, Science and Culture.
The code OXBASH in the program library of Computer Center,
University of Tokyo is used.
It is acknowledged that VAX6440, in Meson Science Laboratory,
University of Tokyo, is also utilized.

\section*{Appendix}
\renewcommand{\theequation}{A\arabic{equation}}
\setcounter{equation}{0}

In this Appendix we review how the tail form
of a bound-state wavefunction
is connected with separation energies.
The Schr\"{o}dinger equation for the nucleus $A$ is
\be H_A \psi(A) = E(A) \psi(A) . \label{eq:SchA} \ee
For the sake of simplicity,
we shall deal only with the ground state of $A$.
It is straightforward to extend this discussion to excited states.
If we consider the breakup channel $A\rightarrow A' + a$,
where $a$ represents a single nucleon or a cluster of nucleons,
it is convenient to rewrite it as
\bea \psi(A) &=& {\cal N} \sum_{i,j} {\cal A}
 [\varphi_{A',a}(\{{\bm R}_\alpha\}) \psi(A';i) \psi(a;j)] ,
 \label{eq:wfAbr} \\
 H_A &=& H_{A'} + H_a + T_R + V_{A',a}(\{{\bm R}_\alpha\}) .
 \label{eq:HAbr} \eea
Here $\varphi_{A',a}(\{{\bm R}_\alpha\})$
denotes the wavefunction
with respect to the relative coordinates
between a nucleon involved in $A'$ and another nucleon in $a$.
The expression $\cal A$ stands for
the antisymmetrizer for all nucleons
and ${\cal N}$ an appropriate normalization constant.
The indices $i$ and $j$ represent various configurations
of the partition $A'$ and $a$.
The operator $T_R$ in the hamiltonian
represents the kinetic energy regarding the variable ${\bm R}$,
where ${\bm R}$ is the relative coordinate
between the center-of-mass of $A'$ and that of $a$,
and $V_{A',a}(\{{\bm R}_\alpha\})$ the interaction
between $A'$ and $a$.

For sufficiently large $R$ in the breakup channel,
we can take $\psi(A';i)$ and $\psi(a;j)$
as eigenstates of $H_{A'}$ and $H_a$, respectively.
Then $\varphi_{A',a}(\{{\bm R}_\alpha\})$ becomes a function
only of ${\bm R}$
and ${\cal A}$ can be regarded as identity.
This will be satisfied when $R$ exceeds the range $R_N$
outside which the nuclear force between $A'$ and $a$ vanishes.
The potential term $V$ consists
of the nuclear part $V^N$ and the Coulomb part $V^C$.
Because of the finite-range nature of the nuclear force,
$V$ is described only by $V^C$ for $R>R_N$.
Hereafter we restrict ourselves to $R>R_N$,
abbreviating $\varphi_{A',a}(\{{\bm R}_\alpha\})$
as $\varphi({\bm R})$.

It is sufficient to consider only the ground states of $A'$ and $a$
for discussing the asymptotic behavior of $\psi(A)$.
Substituting Eqs.(\ref{eq:wfAbr},\ref{eq:HAbr})
into Eq.(\ref{eq:SchA})
and integrating out the internal variables of $A'$ and $a$,
we obtain
\be [T_R+V^C]\varphi({\bm R}) = [E(A)-E(A')-E(a)]\varphi({\bm R}) ,
{}~~\mbox{for $R>R_N$}. \label{eq:Sch-rel} \ee
The separation energy of $A$ for the particle $a$ is defined by
\be S_a (A)= -[E(A)-E(A')-E(a)] . \ee
Therefore we can rewrite Eq.(\ref{eq:Sch-rel}) as
\be \left[-{{\nabla_R^2}\over{2\mu}} + V^C \right]\varphi({\bm R})
 = -S_a\varphi({\bm R}) ,
{}~~\mbox{for $R>R_N$}, \label{eq:asymp} \ee
where the reduced mass for the $A'+a$ system is denoted by $\mu$.
Remark that we are considering a bound-state
of $A$ ({\it i.e.}, $S_a>0$),
leading to the boundary condition
of $\varphi({\bm R})\rightarrow 0$ for $R\rightarrow \infty$.
The strongest damping factor in the asymptotic region is determined
from the following equation (by neglecting terms of $O(R^{-1})$),
\be -{1\over{2\mu}} {{\partial^2}\over{\partial R^2}}
 \varphi({\bm R}) \sim -S_a\varphi({\bm R}) ,
{}~~\mbox{for large $R$}. \ee
Therefore, for extremely large $R$,
the wavefunction $\varphi({\bm R})$ necessarily has
the damping form $f(R) e^{-\sqrt{2\mu S_a}R}$,
where $f(R)$ represents a function without an exponential damping
tail.
Indeed, suppose that $a$ is a neutron, for which $V^C=0$,
Eq.(\ref{eq:asymp}) provides us with the following solution,
\be \varphi_{lm}({\bm R}) \simeq
 \xi' h_l^{(1)}(i\sqrt{2\mu S_a} R) Y^{(l)}_m(\hat{\bm R}) ,
{}~~\mbox{for $R>R_N$}, \ee
where $h_l^{(1)}(x)$ expresses the spherical Hankel function,
$\hat{\bm R}$ indicates ${\bm R}/R$
and $\xi'$ denotes an amplitude.
This immediately leads to
the asymptotic form of $\varphi({\bm R})$ as
\be \varphi_{lm}({\bm R}) \simeq \xi
 {{e^{- \sqrt{2\mu S_a} R}}\over{R}} Y^{(l)}_m(\hat{\bm R}) ,
{}~~\mbox{for $R>R_N$ and $R\gg 1/\sqrt{2\mu S_a}$}.
 \label{eq:asymp-n} \ee
The asymptotic forms of other channels are obtained
in a similar manner.
It should be kept in mind that
the value of the amplitude $\xi$ in Eq.(\ref{eq:asymp-n})
cannot be determined within this asymptotic treatment.
It is pointed out that the channel
with the smallest separation energy
yields the farthest reaching tail of the total wavefunction of $A$.
Thus the wavefunction of a bound ground-state
has a definite tail form given
by the corresponding separation energy.

\clearpage
\pagestyle{empty}
\begin{table}
\caption{$Q$-moments ($e{\rm fm}^2$)
and $B(E2)$ values ($e^2{\rm fm}^4$)
in $A=6$--$10$ nuclei.~~~~~~~~~~\label{tab:E2}}
\begin{tabular}{|c|c||r|r||r@{$\pm$}r||r|}
   \hline
    nucleus & quantity & \multicolumn{2}{c||}{$(0+2)\hbar\omega$}
     & \multicolumn{2}{c||}{exp.}
     & \multicolumn{1}{c|}{$0\hbar\omega$}\\
    \cline{3-4}
    && \multicolumn{1}{c|}{cal.(A)} & \multicolumn{1}{c||}{cal.(B)}
     & \multicolumn{2}{c||}{} & \multicolumn{1}{c|}{cal.(C)}\\
   \hline
     $^6$Li & $Q(1_1^+)$ & $0.09$ & $0.10$ & $-0.08$&$0.01^{a~}$
       & $-1.83$ \\
      & $B(E2; 3_1^+ \rightarrow 1_1^+)$ & $5.95$ & $7.21$
       & $10.7~$&$0.8~^{b~}$ & $8.13$ \\
      & $B(E2; 2_1^+ \rightarrow 1_1^+)$ & $9.17$ & $11.10$
       & $4.4~$&$2.3~^{b~}$ & $3.93$ \\
     $^7$Li & $Q([{3\over 2}]_1^-)$ & $-3.79$ & $-4.29$
      & $-4.06$&$^{a~}$ & $-4.86$ \\
      & $B(E2; [{1\over 2}]_1^- \rightarrow [{3\over 2}]_1^-)$
       & $13.21$ & $17.21$ & $15.7~$&$1.0~^{b~}$ & $22.74$ \\
      & $B(E2; [{7\over 2}]_1^- \rightarrow [{3\over 2}]_1^-)$
       & $5.96$ & $7.83$ & $3.42$&$^{b~}$ & $8.53$ \\
     $^8$Li & $Q(2_1^+)$ & $2.78$ & $3.21$ & $3.15$&$0.05^{a~}$
       & $3.24$ \\
      & $B(E2; 1_1^+ \rightarrow 2_1^+)$ & $3.94$ & $5.30$
       & $75~~~$&$17~~~^{d~}$ & $6.72$ \\
     $^8$B & $Q(2_1^+)$ & $5.84$ & $6.27$ & $6.83$&$0.21^{c\dag}$
      & $5.17$ \\
     $^9$Li & $Q([{3\over 2}]_1^-)$ & $-3.89$ & $-4.36$
      & $-3.6~$&$0.7~^{a\dag}$ & $-5.05$ \\
     $^9$Be & $Q([{3\over 2}]_1^-)$ & $5.46$ & $5.98$
       & $5.3~$&$0.3~^{a~}$ & $5.36$ \\
      & $B(E2; [{5\over 2}]_1^- \rightarrow [{3\over 2}]_1^-)$
       & $26.39$ & $32.02$ & $27.1~$&$2.0~^{b~}$ & $31.94$ \\
      & $B(E2; [{7\over 2}]_1^- \rightarrow [{3\over 2}]_1^-)$
       & $9.75$ & $11.63$ & $7.0~$&$3.0~^{b~}$ & $10.74$ \\
     $^{10}$Be & $B(E2; 2_1^+ \rightarrow 0_1^+)$ & $13.48$
       & $16.26$ & $10.2~$&$1.0~^{b~}$ & $17.38$ \\
      & $B(E2; 0_2^+ \rightarrow 2_1^+)$ & $5.87$ & $7.20$
       & $3.2~$&$1.9~^{b~}$ & $0.01$ \\
     $^{10}$B & $Q(3_1^+)$ & $9.62$ & $10.58$ & $8.47$&$0.06^{a~}$
       & $10.70$ \\
      & $B(E2; 1_1^+ \rightarrow 3_1^+)$ & $1.12$ & $1.35$
       & $4.13$&$0.06^{b~}$ & $9.76$ \\
      & $B(E2; 1_2^+ \rightarrow 3_1^+)$ & $7.11$ & $8.61$
       & $1.71$&$0.26^{b~}$ & $1.35$ \\
      & $B(E2; 1_2^+ \rightarrow 1_1^+)$ & $3.21$ & $3.88$
       & $0.83$&$0.40^{b~}$ & $2.03$ \\
      & $B(E2; 3_2^+ \rightarrow 1_1^+)$ & $13.86$ & $16.77$
       & $20.5~$&$2.6~^{b~}$ & $9.68$ \\
     $^{10}$C & $B(E2; 2_1^+ \rightarrow 0_1^+)$ & $12.54$ & $15.22$
       & $12.3~$&$2.1~^{b~}$ & $15.01$ \\
   \hline
\end{tabular}
\\
cal.(A): $e_\pi^{\rm eff}=e$, $e_\nu^{\rm eff}=0$.\\
cal.(B): $e_\pi^{\rm eff}=1.05e$, $e_\nu^{\rm eff}=0.05e$.\\
cal.(C): $e_\pi^{\rm eff}=1.5e$, $e_\nu^{\rm eff}=0.5e$
($b=1.77{\rm fm}$).\\
$^a$) Ref.\cite{ref:moment};
$^b$) Ref.\cite{ref:Selove};
$^c$) Ref.\cite{ref:Minamisono};
$^d$) Ref.\cite{ref:Li8E2}.\\
$^{\dag}$) The sign is speculated from the $0\hbar\omega$
and $(0+2)\hbar\omega$ shell-model calculations.\\
\end{table}

\clearpage
\section*{Figure Captions}
\begin{list}{Fig.\thefig:\hfill}{\usecounter{fig}}
\item Energy levels of $^8$Li and $^8$B.
Coulomb energies are subtracted
in the same way as Ref.\cite{ref:Wolters}.
The levels labeled `Utr.' are obtained
by the $(0+2)\hbar\omega$ calculation
with the interaction of Wolters {\it et al.},
while those labeled `CK' are by the $0\hbar\omega$ calculation
with the Cohen-Kurath interaction.
\label{fig:energy}
\item $R_T$ values defined in Eq.(\ref{eq:R})
for $Q$-moments and $E2$ transitions.
They are displayed according to the sequence
of moment or transition probability in Table~\ref{tab:E2},
except for $B(E2; 0_2^+ \rightarrow 2_1^+)$ of $^{10}$Be.
Each $R_{T=1}$ value concerns the same transition (or $Q$-moment)
as the $R_{T=0}$ value shown right above.
See the text for details.
\label{fig:effch}
\end{list}

\end{document}